%% file: rnaas.tex
\documentclass[RNAAS]{aastex62}

\begin{document}

\title{Some Bright Stars with Smooth Continua for Calibrating the Response of High Resolution Spectrographs}


\correspondingauthor{Jason T.\ Wright}
\email{astrowright@gmail.com}

\author{Kelsey Clubb}
\affiliation{Department of Astronomy, 601 Campbell Hall, University of California, Berkeley, CA 94720-3411, USA}

\author[0000-0001-8638-0320]{Andrew W.\ Howard}
\affiliation{California Institute of Technology, 1200 E California Boulevard, Pasadena, CA, 91125, USA}

\author[0000-0002-0531-1073]{Howard Isaacson}
\affiliation{Department of Astronomy, 601 Campbell Hall, University of California, Berkeley, CA 94720-3411, USA}

\author[0000-0002-2909-0113]{Geoffrey W.\ Marcy}
\affiliation{Department of Astronomy, 601 Campbell Hall, University of California, Berkeley, CA 94720-3411, USA}

\author[0000-0001-6160-5888]{Jason T.\ Wright}
\affiliation{Department of Astronomy \& Astrophysics, 525 Davey Laboratory,
The Pennsylvania State University,
University Park, PA, 16802, USA}
\affiliation{Center for Exoplanets and Habitable Worlds, 525 Davey Laboratory,
The Pennsylvania State University,
University Park, PA, 16802, USA}




\section{} 

When characterizing a high resolution echelle spectrograph, for instance for precise Doppler work, it is useful to observe featureless sources such as quartz lamps or hot stars to determine the response of the instrument.  Such sources provide a way to determine the blaze function of the orders, pixel-to-pixel variations in the detector, fringing in the system, and other important characteristics.

In some cases it is especially important that the source be astronomical, since its light then follows the same optical path and lands on the same detector pixels as the science light. Hot stars also provide significant flux in the blue, where many quartz lamps are not bright.

It is important that such astronomical sources be as spectrally featureless as possible, and bright enough that observation of them does not consume too much time. In principle, B and early A dwarfs, subgiants, and giants should be ideal: their high luminosity means many have $V<7$, they are common enough to be found in all parts of the sky, and their characteristic rotational broadening washes out all shallow spectral features, leaving only the Balmer lines (in the optical) which require some care but occupy only a small portion of the spectrum.

In practice, however, many B or early A stars do not provide a smooth continuum, whether because they are not rotating rapidly enough or for some other reason. In fact, we have found that published rotational velocities and temperatures are not a specific and sensitive guide to whether a star's continuum will be smooth. A useful resource for observers, therefore, is a list of ``good'' hot stars: bright, blue stars known empirically to have no lines or other spectral features beyond the Balmer series with minima below 95\% of the continuum.

We have compiled a list of bright, early-type stars visible from Northern Hemisphere telescopes. This list includes all stars listed in the Yale Bright Star Catalog \citep{HRCatalog} as being single with $V<5.5$, $v\sin{i}> 175$ km s$^{-1}$, and declination $>-30$, and many other  hot stars that we have found useful for calibration purposes. 

From this list, one of us (K.\ C.)  visually inspected the entire ``iodine region'' (4950--6200\AA) of each hot star's spectrum taken at Keck/HIRES \citep{Vogt94} as part of the California Planet Survey \citep{Howard10a} and its predecessors.  We categorize each star as being a ``good'' or ``bad'' target for calibration based on whether it has metal lines with depths below 95\% of the continuum level.

In Table~\ref{table} (which is a stub for the machine readable version provided online with this Research Note) we provide each star's HR number, coordinates, V magnitude,  spectral type, and rotational broadening from the Bright Star Catalog; whether it is a ``good'' or ``bad'' target according to our vetting; and our observing notes regarding binarity. We also list many stars for which we do not have spectra (marked neither ``good'' nor ``bad'') so others may continue the vetting process for those.

The list here of ``bad'' stars may also be of interest in studies of hot, slowly rotating stars.

\begin{deluxetable}{rrrrlcrr}
\tablehead{\colhead{Star Name} & \colhead{RA} & \colhead{Dec} & \colhead{V} & \colhead{SpType} & \colhead{$v\sin{i}$} & \colhead{good?} & \colhead{Notes} \\
&\multicolumn{2}{c}{Equinox 2000}&&&km s$^{-1}$&&}
\tablecaption{Sample of the table to illustrate its contents. The full table is only available in machine readable format. The first six columns are from the Yale Bright Star Catalog. \label{table}}
\startdata
\input{Bstar_final.tex}
\enddata
\end{deluxetable}

\acknowledgments

We thank the many observers of the California Planet Survey for their work collecting the data that makes this Research Note possible.

This research has made use of the SIMBAD database, operated at CDS, Strasbourg, France, and NASA's Astrophysics Data System Bibliographic Services.

\bibliography{references}

\end{document}

%% file: Bstar_final.tex
    HR9098 & 00 03 44.4 & -17 20 10 & 4.55  &   B9V  &  186 &  good   &                             \\
     HR179 & 00 42 03.9 & +50 30 45 & 4.80  &   B2V  &  211 &  good   &                             \\
     HR193 & 00 44 43.5 & +48 17 04 & 4.54  & B5III  &  260 &  bad    &                             \\
     HR264 & 00 56 42.5 & +60 43 00 & 2.47  &  B0IV  &  300 &         &                             \\
     HR545 & 01 53 31.8 & +19 17 45 & 4.83  &   B9V  &  152 &  good   &  northern component 7$\arcsec$      